\documentclass[amsmath,superscriptaddress,preprint]{revtex4}
\usepackage{graphicx}
\usepackage{dcolumn}
\usepackage{bm}
\usepackage{amssymb}
\usepackage{ulem}
\usepackage{color}
\usepackage{amsfonts}

\begin{document}
\title{Phonon thermal Hall effect in a metallic spin ice}
\author{Taiki Uehara}
\affiliation{Department of Physics, Gakushuin University, Tokyo 171-8588, Japan.}

\author{Takumi Ohtsuki}
\affiliation{Institute for Solid State Physics, The University of Tokyo, Kashiwa 277-8581, Japan.}

\author{Masafumi~Udagawa}
\affiliation{Department of Physics, Gakushuin University, Tokyo 171-8588, Japan.}

\author{Satoru Nakatsuji}
\affiliation{Institute for Solid State Physics, The University of Tokyo, Kashiwa 277-8581, Japan.}
\affiliation{Department of Physics, The University of Tokyo, Tokyo 113-0033, Japan.}
\affiliation{The Institute for Quantum Matter and the Department of Physics and Astronomy,
The Johns Hopkins University, Baltimore, MD 21218, USA.}

\author{Yo Machida}
\affiliation{Department of Physics, Gakushuin University, Tokyo 171-8588, Japan.}

\date{\today}

\begin{abstract}
It has become common knowledge that phonons can generate thermal Hall effect in a wide variety of materials, although the underlying mechanism is still controversial. We study longitudinal $\kappa_{xx}$ and transverse $\kappa_{xy}$ thermal conductivity in Pr$_2$Ir$_2$O$_7$, which is a metallic analogue of spin ice. 
Despite the presence of mobile charge carriers, we find that both $\kappa_{xx}$ and $\kappa_{xy}$ are dominated by phonons.
A $T/H$ scaling of $\kappa_{xx}$ unambiguously reveals that longitudinal heat current is substantially impeded by resonant scattering of phonons on paramagnetic spins. Upon cooling, the resonant scattering is strongly affected by a development of spin ice correlation 
and $\kappa_{xx}$ deviates from the scaling in an anisotropic way with respect to field directions.
Strikingly, a set of the $\kappa_{xx}$ and $\kappa_{xy}$ data clearly shows that $\kappa_{xy}$ correlates with $\kappa_{xx}$ in its response to magnetic field including a success of the $T/H$ scaling and its failure at low temperature.
This remarkable correlation provides solid evidence that an indispensable role is played by spin-phonon scattering not only for hindering the longitudinal heat conduction, but also for generating the transverse response. 
\end{abstract}
\maketitle

When heat current carried by electrons is subject to magnetic field applied normal to the current, a trajectory of electrons is curved by the Lorentz force and a transverse temperature gradient is developed in the direction both perpendicular to the heat current and the magnetic field. This phenomenon, dubbed the thermal Hall effect, has been believed to be restricted to materials in which there are mobile charge carriers. However, it is shown that even if the carriers of heat are neutral, the thermal Hall effect arises in several materials including magnetic insulators~\cite{strohm,chen}, multiferroics~\cite{ideue}, spin liquid candidates~\cite{hirschberger,sugii,kasahara1,kasahara2,hentrich,akazawa,yokoi,lefrancois}, Mott insulators~\cite{grissonnanche1,grissonnanche2,boulanger}, and nonmagnetic insulator~\cite{STO}, providing a new insight on heat transport in solids. 
In some of the preceding materials, phonons are identified as the heat carriers responsible for the thermal Hall effect~\cite{strohm,chen,ideue,sugii,akazawa,lefrancois,grissonnanche1,grissonnanche2,boulanger,STO}.
Despite growing number of reports presenting the phonon Hall effect, little is known about the microscopic mechanism~\cite{sheng,jswang,zhang,qin,mori,saito,yang,flebus,sun}.

To address this issue, we carried out measurements of thermal conductivity tensors in Pr$_2$Ir$_2$O$_7$, which is a kind of `treasure trove' of attractive physical properties including Kondo effect in a frustrated magnet~\cite{nakatsuji}, topological Hall effect~\cite{machida1,machida2}, spin ice state in a metal~\cite{machida2}, quantum criticality~\cite{tokiwapr}, and Luttinger semimetal with a quadratic band touching~\cite{kondo}. Among them, the relevant features to this study are absence of a long-range magnetic order down to the lowest temperature measured and semimetallicity with low carrier density.
The former prevents contamination of magnon contribution in the heat transport coefficients. While the metallicity enables a precise estimation of electron contribution via the Wiedemann-Franz law, the low density of electron carriers leaves room for detection of the thermal Hall effect by chargeless carriers at the same time.

We show that longitudinal thermal conductivity $\kappa_{xx}$, in which phonon contribution by far dominates electron contribution, is largely degraded by spin-phonon scattering as low as that of amorphous silica. 
$\kappa_{xx}$ is further lowered by magnetic field due to resonant scattering between phonons and paramagnetic spins as evidenced by a $T/H$ scaling. Upon cooling, magnetic fluctuations arising from spin ice correlation affects the resonant scattering by rendering another source of local level splitting through an exchange field.

An anisotropic suppression of the fluctuations with respect to magnetic field directions gives rise to anisotropic enhancement of $\kappa_{xx}$, and concomitantly a deviation from the $T/H$ scaling. 
Despite the presence of mobile electrons, we detected finite thermal Hall conductivity $\kappa_{xy}$ mostly generated by phonons. Unexpectedly, we find striking similarities between $\kappa_{xx}$ and $\kappa_{xy}$ in their response to magnetic field, but importantly $\kappa_{xy}$ behaves oppositely to $\kappa_{xx}$. Namely, $\kappa_{xy}$ increases when $\kappa_{xx}$ is suppressed by the resonant phonon scattering.  Conversely, $\kappa_{xy}$ decreases when $\kappa_{xx}$ is enhanced due to the suppression of magnetic fluctuations by field. This observation explicitly indicates that a single mechanism drives both longitudinal and transverse thermal response, and spin-phonon coupling which affects the mean-free-path of phonons has a skew scattering component.

Single crystals of Pr$_2$Ir$_2$O$_7$ were grown by a flux method~\cite{millican}. 
Two different samples were used for the thermal transport measurements under magnetic field applied parallel to the [111] and [001] directions.
Longitudinal thermal conductivity $\kappa_{xx}$ and thermal Hall conductivity $\kappa_{xy}$ were measured by the standard steady state method in high vacuum. 

Figure 1 shows temperature dependence of longitudinal thermal conductivity $\kappa_{xx}$ of Pr$_2$Ir$_2$O$_7$ measured under zero field by applying the heat current $Q$ parallel to the (001) plane ($Q\parallel(001)$). The data is shown together with those of insulating pyrochlore magnets, Yb$_2$Ti$_2$O$_7$~\cite{tokiwa}, Y$_2$Ti$_2$O$_7$~\cite{kolland}, Dy$_2$Ti$_2$O$_7$~\cite{kolland}, and Tb$_2$Ti$_2$O$_7$~\cite{li} where heat conduction is dominated by phonons.
As shown in the inset of Fig.~1(a), electronic contribution $L_0\sigma_{xx}T$ to $\kappa_{xx}$ estimated using the Wiedemann-Franz (WF) law is more than one order of magnitude smaller than $\kappa_{xx}$ for both $Q\parallel(001)$ and $Q\parallel(111)$, which indicates that heat is predominantly transported by phonons. 
Interestingly, the magnitude of $\kappa_{xx}$ is extremely small, and approaches that of amorphous silica~\cite{zeller}.
Moreover, the so-called phonon peak, which is characteristic to phononic thermal conductivity in insulating crystalline solids, is absent.
It is shown that the structural disorder has negligible effect on the Raman phonon spectra in the sample from the same source~\cite{xu}. We thus stress that the low $\kappa_{xx}$ is not due to phonon scattering by random disorder.
Since a position of the phonon peak is scaled by the Debye temperature, which is typically around 300-400 K for pyrochlore oxides~\cite{subramanian}, $\kappa_{xx}$ for Yb$_2$Ti$_2$O$_7$, Y$_2$Ti$_2$O$_7$ and Dy$_2$Ti$_2$O$_7$ has the peaks at similar temperature ($\sim$ 10 K). At high temperatures exceeding the peak, the magnitude of thermal conductivity is set by rate of collisions between thermally excited phonons whose number is also scaled by the Debye temperature. Thus, it is quite reasonable that $\kappa_{xx}$ for Yb$_2$Ti$_2$O$_7$ and Dy$_2$Ti$_2$O$_7$ are close to each other at high temperatures. 

By contrast, in Pr$_2$Ir$_2$O$_7$ the phonon peak is absent and $\kappa_{xx}$ is smaller than those of Y$_2$Ti$_2$O$_7$ and Dy$_2$Ti$_2$O$_7$ by a factor of five even at high temperatures, although our samples are the crystalline solids and the Debye temperature of 400 K~\cite{ghosh} is similar to the other pyrochlore oxides. This suggests the presence of additional scatterers of phonons except for other phonons and disorder. Notably, thermal conductivity is also small and the phonon peak is absent in spin liquid candidate Tb$_2$Ti$_2$O$_7$ (Fig. 1(a)) where these striking features are attributed to strong phonon scattering by magnetic fluctuations~\cite{li}. 
As we see below, spin-phonon scattering is a
leading mechanism of the low phonon thermal conductivity in Pr$_2$Ir$_2$O$_7$.
An intrinsic scattering of phonon by mobile electrons may be additional thermal impedance of the heat flow.
 
In Figs. 1(b) and (c), magneto-thermal conductivity $\{\kappa_{xx}(H)-\kappa_{xx}(0)\}/\kappa_{xx}(0)$ are shown for $H\parallel[111]$ and $H\parallel[001]$, respectively. For both directions, $\kappa_{xx}$ first decreases with field and takes a minimum. 
On warming, a position of minimum defined as $H_{\rm min}$ shifts to higher fields. (See the inset of Fig.1(c) for systematic change of $H_{\rm min}$ with temperature for $H\parallel[001]$.) 
As shown in the inset of Fig. 1(b), $H_{\rm min}$ increases linearly with temperature, $H_{\rm min}\sim T$, regardless with the field directions. 
Such a behavior has been observed in various paramagnets and is attributed to resonant scattering between phonons and paramagnetic spins~\cite{berman}. The resonance can occur in the presence of a strong spin-phonon coupling when the two-level spin systems split by the Zeeman energy absorb phonon and subsequently emit another phonon of the same energy in an unrelated direction. This spin-flip process effectively scatters phonons. The scattering becomes the largest when the Zeeman splitting $\Delta E\sim 2M\mu_BH$ ($M$ is magnetization) is equal to phonon energy whose spectrum has a broad maximum at $\sim4k_{\rm B}T$. This causes the minimum in $\kappa_{xx}(H)$ at $H_{\rm min}\sim 2k_{\rm B}T/M\mu_B$ with $H_{\rm min}$ proportional to $T$.
Therefore, field-induced change in the longitudinal thermal conductivity measured at various temperatures is expected to be scaled as a function of $H/T$ with a minimum at $H_{\rm min}/T\sim2k_{\rm B}/M\mu_B$.
Such a scaling is demonstrated in Figs. 4(c) and 4(d) where $\Delta\kappa_{xx}(H)=\kappa_{xx}(H)-\kappa_{xx}(0)$ normalized by its minimum value $\Delta\kappa_{xx}(H)/\kappa_{xx}^{min}$ is plotted against $H/T$ for $H\parallel[111]$ and $H\parallel[001]$, respectively. Remarkably, all data fall onto the same curve except for $H/T>1$ and present minimum at $H/T\sim$ 1.
This result unambiguously indicates that $\kappa_{xx}(H)$ is controlled by the resonant phonon scattering in the region of $H<T$. 
For the free Pr$^{3+}$ ion, the magnetization is expected to be $M=g_JJ=3.2$, where $g_J$ and $J$ represent the Land$\acute{\rm e}$'s $g$ factor and the total angular moment, which gives $H_{\rm min}/T\sim2k_{\rm B}/3.2\mu_B\sim0.93$, in good agreement with our observations.
By closer looking at the data, however, one notices that the minimum position is slightly different with respect to the field directions: $H_{\rm min}/T\sim1.25$ and 0.75 for $H\parallel[111]$ and $H\parallel[001]$, respectively.
We will come back to this point later.

Let us turn our attention to the high field regions. With increasing field, $\{\kappa_{xx}(H)-\kappa_{xx}(0)\}/\kappa_{xx}(0)$ becomes positive (Figs.~1(b) and 1(c)) and the $H/T$ scaling becomes failed (Figs.~4(c) and 4(d)). Concomitantly, we resolved a clear anisotropy in $\{\kappa_{xx}(H)-\kappa_{xx}(0)\}/\kappa_{xx}(0)$ especially at low temperatures: while $\{\kappa_{xx}(H)-\kappa_{xx}(0)\}/\kappa_{xx}(0)$ for $H\parallel[111]$ increases with a concave curvature, the one for $H\parallel[001]$ increases with a convex curvature and shows a tendency to saturate at low temperatures. 
Since the resonant scattering between phonons and paramagnetic spins is responsible for the negative magneto-thermal conductivity, the observed anisotropic recovery of $\kappa_{xx}(H)$ implies the magnitude of resonant scattering is substantially influenced by spin correlation.

In magnetic materials, magnetic fluctuations yield strong scattering on phonons and significantly suppress phononic heat conduction~\cite{sharma}. An application of magnetic field, however, weakens magnetic fluctuations and leads to a striking enhancement of phonon thermal conductivity~\cite{wang,tokiwa,li}. The observed response of $\kappa_{xx}$ to magnetic fields can be understood based on this line of thought. In particular, anisotropic magneto-thermal conductivity explicitly indicates a vital role of phonon scattering by fluctuating spins with spin ice correlation. In spin ice state~\cite{SpinIce}, the spin system fluctuates between the energetically equivalent ``2-in, 2-out'' configurations within the ground state manifold. This gives rise to strong magnetic fluctuations. 
The macroscopic degeneracy is lifted by the external magnetic field in an anisotropic way~\cite{fukazawa,hiroi}.
Magnetic field along the [001] direction steeply lift the ground state degeneracy and suppress the magnetic fluctuations because the stable spin configuration is uniquely determined as one of the six equivalent ``2-in, 2-out'' configurations by the field. For $H\parallel[111]$, ``3-in, 1-out/1-in, 3-out'' configuration is energetically favored in high field limit. However, due to a smaller Zeeman energy gain for the spins on the Kagome plane with the ``3-in, 1-out/1-in, 3-out'' configuration than the ``2-in, 2-out'' configuration, the system remains in spin ice manifold and preserves the strong magnetic fluctuations up to higher field~\cite{UdagawaOgataHiroi,MoessnerSondhi}.

This anisotropic suppression of magnetic fluctuations brings about positive and anisotropic magneto-thermal conductivity. For $H\parallel[001]$, the steep suppression of the magnetic fluctuations yields the rapid rise of $\{\kappa_{xx}(H)-\kappa_{xx}(0)\}/\kappa_{xx}(0)$ (Fig. 1(c)). Once the polarized state with the ``2-in, 2-out'' configuration is stabilized by the fields and fluctuations are totally suppressed, $\kappa_{xx}$ gets saturated to a value which is purely dominated by phonons. Namely, the resonant scattering does not work any more, since the spins are fully polarized in a saturation field, and the number of phonons carrying sufficient energy to flip the spin is exponentially suppressed.
Meanwhile, the persistence of the magnetic fluctuations for $H\parallel[111]$ yields the slower rise of $\{\kappa_{xx}(H)-\kappa_{xx}(0)\}/\kappa_{xx}(0)$ (Fig. 1(b)).

Figure~1(d) shows temperature dependence of $\{\kappa_{xx}(H)-\kappa_{xx}(0)\}/\kappa_{xx}(0)$ measured at 9 T for $H\parallel[111]$ and $H\parallel[001]$. On cooling, $\{\kappa_{xx}(H)-\kappa_{xx}(0)\}/\kappa_{xx}(0)$ changes a sign from negative to positive around 4 K and 7 K for $H\parallel[111]$ and $H\parallel[001]$, respectively. As mentioned above, the resonant phonon scattering is strongly influenced by spin correlation through a local exchange field. However, by applying a large magnetic field, or equivalently at low temperatures, the local spin splitting becomes mainly determined by external magnetic field. This crossover causes a gradual change from the negative to positive magneto-thermal conductivity.
In that sense, the sign-change temperature can be regarded as a lower bound of onset temperature below which the spin-ice correlation sets in. Notably, this temperature roughly coincides with a resistivity minimum (see the inset of Fig.~1(d)) which is another consequence of the spin-ice correlation while in this case the correlated spins interact with conduction electrons~\cite{udagawa}.

Now, let us discuss an implication of the spin-ice correlations to the anisotropy in $H_{\rm min}$.
Under the spin ice state, the Zeeman splitting $\Delta E\sim 2M\mu_BH$ of the ground state doublet is anisotropic with respect to the field directions due to anisotropy in magnetization $M$~\cite{nakatsuji}. Accordingly, given the relation of $H_{\rm min}\sim 2k_{\rm B}T/M\mu_B$, $H_{\rm min}$ is anisotropic and its anisotropic ratio between [111] and [001] directions is expected to be held a relation of $H_{\rm min}^{\rm [111]}/H_{\rm min}^{\rm [001]}\sim M_{\rm [001]}/M_{\rm [111]}$. This means that at a given temperature the larger Zeeman splitting $\Delta E$ due to the larger $M$ satisfies the condition of resonance at the lower field. In fact, the anisotropic ratio of $H_{\rm min}$, $(H_{\rm min}^{\rm [111]}/T)/(H_{\rm min}^{\rm [001]}/T)=1.25/0.75\sim$ 1.67 extracted from Figs.~4(c) and 4(d), is in good agreement with magnetization anisotropy expected for the ``2-in, 2-out'' configuration, $M_{\rm [001]}/M_{\rm [111]}=\{g_JJ(1/\sqrt{3})\}/\{g_JJ(1+1/3\times1)/4\}\sim$ 1.73. 

Our argument that the spin-phonon scattering controls the evolution of $\kappa_{xx}(H)$ is supported by a theoretical calculation.
We model the interaction between the Pr doublets by a simple spin-ice-type Ising model. Moreover we consider a transverse coupling between the Pr doublets and acoustic phonons, as naturally expected from the non-Kramers nature of Pr doublets~\cite{OnodaTanaka,JeffGingras}.
Even within this simple model, we can qualitatively reproduce main experimental features of magneto-thermal transport as shown in Fig.~1(e); the initial negative magneto-thermal conductivity, the presence of minimum, and the positive increase with the convex curvature at the low temperature and high field region.

Having established the dominant role of spin-phonon scattering in the longitudinal thermal conductivity, let us focus on thermal Hall effect.
Temperature dependence of thermal Hall conductivity divided by temperature $\kappa_{xy}/T$ measured under magnetic field of 9 T for $H\parallel[111]$ and $H\parallel[001]$ are shown in Figs. 2(a) and 2(b), respectively. In the same figures, we also show the electronic contribution $L_0\sigma_{xy}$ (left axis) and $\kappa_{xx}/T$ (right axis).
Surprisingly, a sign of $L_0\sigma_{xy}$ is opposite to $\kappa_{xy}/T$ in the whole measured temperature range for $H\parallel[111]$ and $T>4$ K for $H\parallel[001]$, indicating that the thermal Hall effect is mostly governed by carriers except for electrons. 
Moreover, both $\kappa_{xy}/T$ and $\kappa_{xx}/T$ peak around 20-30 K where phonons dominate the longitudinal thermal conductivity because electron contribution accounts only $L_0\sigma_{xy}T/\kappa_{xx}\sim$ 0.6 \% of the total $\kappa_{xx}$ (see the inset of Fig.1(a)). Such a coincidence of peaks in $\kappa_{xx}$ and $\kappa_{xy}$ has been observed in several insulating solids and regarded as a clue to identify the thermal Hall signal generated by phonons~\cite{STO}. 
This result further supports the conjecture that thermal Hall current is carried by phonons in Pr$_2$Ir$_2$O$_7$.
We note a ratio $\kappa_{xy}/\kappa_{xx}\simeq0.4-0.8\times10^{-3}$ around the peak is comparable to that found in materials where phonons have been argued to cause the Hall effect~\cite{chen,lefrancois}.

At temperatures above the peak, the magnitude of $\kappa_{xy}/T$ is comparable with that of Tb$_2$Ti$_2$O$_7$~\cite{hirschberger} and smaller than the unexpectedly large thermal Hall conductivity of SrTiO$_3$~\cite{STO} and La$_2$Cu$_4$O~\cite{grissonnanche1} by a factor of ten (the inset of Fig. 2(b)). 
Below the peak, $\kappa_{xy}/T$ steeply decreases faster than $\kappa_{xx}/T$. $\kappa_{xy}/T$ for $H\parallel[001]$ seems to approach the value expected from the WF law followed by a sign change around 4 K, showing that phonons cease to contribute the Hall response at low temperatures. 

In Figure 2(c) and 2(d), we show magnetic field dependence of thermal Hall conductivity $\kappa_{xy}(H)$ (triangles) together with electron contribution $L_0\sigma_{xy}T(H)$ (dotted lines) estimated by using the WF law for $H\parallel[111]$ and $H\parallel[001]$, respectively.  Again, a sign of $L_0\sigma_{xy}T(H)$ is opposite to $\kappa_{xy}$. By subtracting $L_0\sigma_{xy}T(H)$ from $\kappa_{xy}(H)$, we evaluated thermal Hall conductivity generated by phonons as $\kappa_{xy}^{ph}=\kappa_{xy}-L_0\sigma_{xy}T$ (circles). As seen from the figures, at 20 K $\kappa_{xy}(H)$ increases linearly with $H$ and there is negligible electron contribution for both directions. On cooling, $\kappa_{xy}(H)$ becomes non-monotonic. Namely, $\kappa_{xy}(H)$ shows a peak and subsequently decreases with field. Moreover, a fraction of the electron contribution to $\kappa_{xy}$ slightly increases, which is maximized up to $|L_0\sigma_{xy}T|/\kappa_{xy}\sim$ 28 \% at 2.9 K and $H$ = 4 T $\parallel[111]$. Since the peak remains in $\kappa_{xy}^{ph}(H)$ even after the subtraction of electron contribution, phonons are responsible for the non-monotonic behavior. By further decreasing temperature, anisotropic field response emerges at high fields. $\kappa_{xy}(H)$ for $H\parallel[001]$ is considerably suppressed above its peak field and approaches a value expected from the WF law within an experimental error (the inset of Fig. 2(d)), consistent with what we saw in temperature variation of $\kappa_{xy}/T$ (Fig. 2(b)). By contrast, the suppression is weak for $H\parallel[111]$ and $\kappa_{xy}^{ph}(H)$ remains positive up to 9 T. 

One of the most striking findings of this work is a correlation between field-induced change in $\kappa_{xy}^{ph}$ and $\kappa_{xx}$, which are displayed in the upper and lower panels of Fig. 3, respectively. In each panel, we compare two data taken at the (nearly) same temperature for $H\parallel[111]$ (open circles) and for $H\parallel[001]$ (closed circles). 
In Fig. 3, there are several things of interest. i) The maximum and the minimum appear at nearly the same field in $\kappa_{xy}^{ph}(H)$ and $\Delta\kappa_{xx}(H)$, respectively, and the extreme positions shift to lower field with decreasing temperature. ii) Above 7 K, the relationship in magnitude of $\Delta\kappa_{xx}(H)$ between the two directions is the same as $\kappa_{xy}^{ph}(H)$: the large negative magneto-thermal conductivity is accompanied by the large thermal Hall signal for $H\parallel[111]$, and vice versa for $H\parallel[001]$. iii) Below 5.1 K, steeper $\Delta\kappa_{xx}(H)$ rises above its minimum field, stronger the suppression of $\kappa_{xy}^{ph}(H)$ becomes above its maximum field. Upon cooling, this correlation becomes more significant for $H\parallel[001]$.

From the observation i), the striking resemblance between $\kappa_{xy}^{ph}(H)$ and $\Delta\kappa_{xx}(H)$ led us to expect that the $H/T$ scaling is also valid for $\kappa_{xy}^{ph}(H)$. As demonstrated in Fig. 4(a) and 4(b), the data of $\kappa_{xy}^{ph}(H)$ divided by its maximum value indeed collapse on a single curve for both directions for $H/T<1$ as in the case of $\kappa_{xx}(H)$ (Fig. 4(c) and 4(d)). Moreover, the observation ii) indicates that the strong paramagnetic scattering of phonons that gives rise to the negative magneto-thermal conductivity is an ingredient to enhance the phonon thermal Hall effect. These results provide compelling evidence that a prominent role is played by resonant phonon scattering not only in degrading longitudinal phonon heat conduction, but also in generating the transverse signal.

We note that the scaling of $\kappa_{xy}^{ph}$ also becomes failed for $H/T>1$ (Fig. 4(a) and 4(b)), indicating that the paramagnetic scattering no longer play a major role in this regime. 
Instead, from the remarkable correlation in the observation iii), it is quite natural to identify another source of asymmetric scattering of phonons as magnetic fluctuations. 
Whereas the survival of magnetic fluctuations along the [111] direction yields the sizable $\kappa_{xy}^{ph}$ even after the paramagnetic scattering dies out,
the strong suppression of magnetic fluctuations along the [001] direction results in the substantial decrease of $\kappa_{xy}$ towards the value purely dominated by electrons.
Thus, it is concluded that whatever the spin state is (whether spins are paramagnetic or correlated), when phonons interact with spins, they are asymmetrically scattered and produce the thermal Hall signal.

To attempt to clarify the intriguing thermal Hall phenomena in Pr$_2$Ir$_2$O$_7$, one should seriously take into account the following two facts. First, the evolution of $\kappa_{xx}$ with field can be thoroughly explained by the way spins scatter phonons.
This indicates an intrinsic coupling of phonons to magnetic environment. 
Second, there is a manifest correlation between $\kappa_{xx}$ and $\kappa_{xy}$ in their evolution with field. 
These two facts impose constraints on possible scenarios that the longitudinal and transverse thermal responses should be understood in a unified way in terms of an intrinsic coupling of phonons to spins with a skew component, and make a possibility of the extrinsic origin like the skew scattering of phonons by superstoichiometric rare earth ions~\cite{mori}, oxygen vacancies~\cite{flebus}, and dynamical defects~\cite{sun} unlikely.

\clearpage

\section*{Methods}
\subsection*{Samples}
We used two different single crystals grown by a flux method for the thermal transport measurements under magnetic field applied parallel to the [111] and [001] directions. The [111] ($H\parallel[111]$) and [001] ($H\parallel[001]$) samples have plate-like shape with dimensions of 1.7(width) $\times$ 2.1(length) mm$^2$ in the (111) plane and 1.1(width) $\times$ 2.1(length) mm$^2$ in the (001) plane, respectively. Thicknesses of the samples are about 0.5 mm. 

\subsection*{Thermal transport measurements}
Longitudinal thermal conductivity $\kappa_{xx}$ and thermal Hall conductivity $\kappa_{xy}$ were measured by the
standard steady state method in high vacuum. The heat flow $Q$ was injected in the (111) and (001) planes for the [111] and [001] samples, respectively, by heating a chip resistor attached to one end of the sample. 
The other end of the sample was attached to an insulating LiF plate, which was used as a cold thermal bath. The longitudinal $\Delta T_x$ and transverse $\Delta T_y$ temperature differences were determined by Cernox thermometers.
The thermometers and the heater were connected by gold wires ($\phi=25\mu$m) and heat-cured silver paint (Dupont 6838) to the sample. The contact resistances were
10 m$\Omega$.
To remove the longitudinal response from the raw data due to misalignment of the contacts, we anti-symmetrized it as $\Delta T_y(H)=\{\Delta T_y(+H)-\Delta T_y(-H)\}/2$.
$\kappa_{xx}$ and $\kappa_{xy}$ were obtained from the longitudinal thermal resistivity, $w_{xx}=(\Delta T_x/Q)(wt/l)$, and the thermal Hall resistivity, $w_{xy}=(\Delta T_y/Q)t$,
as $\kappa_{xx} = w_{xx}/(w_{xx}^2+w_{xy}^2)$ and $\kappa_{xy}=-w_{xy}/(w_{xx}^2+w_{xy}^2)$. Here, $l$, $w$, and $t$ are length between the contacts, width, and thickness of the samples, respectively.
The electrical (Hall) resistivity measurements were done by using the same contacts and gold wires.
$\kappa_{xx}$ and $\kappa_{xy}$ were checked to be independent of the thermal gradient by changing $\Delta T_x/T$ in the range of 1–20 $\%$.
Since $\Delta T_y$ is tiny, which is as small as 0.1 mK, at low temperatures and the scattering of the data is large, the measurements were repeated several times and the data is averaged.
Error bars in the main figures represent standard deviation.

\subsection*{Computational}
Here, we summarize a theoretical formulation to calculate longitudinal thermal conductivity of acoustic phonons, as shown in Fig.1(e). We assume two kinds of scattering centers, non-magnetic impurities and localized Pr moments.
The former gives a scattering rate weakly dependent on the energy of phonons, which results in the normal $\propto T^3$ behavior of phonon thermal conductivity in the low temperature limit.
The latter scattering process is characteristic of this system, in particular, the non-Kramers nature of Pr moments.
It was pointed out that the transverse components of Pr doublets behave as magnetic quadrupole rather than dipole in the Pr pyrochlore oxides~\cite{OnodaTanaka,JeffGingras}.
Consequently, the lattice deformation couples to the transverse components of Pr doublets, or conversely, the acoustic phonons are scattered inelastically through the flip of Pr doublets.

Combining these two types of scattering processes, the thermal conductivity can be concisely written as
\begin{eqnarray}
\kappa_{xx} = \kappa_0\Bigl(1 - \frac{\delta}{T^5}\frac{1}{N}\sum_j\frac{\Delta_j^4}{\sinh^2\frac{\Delta_j}{2T}}\Bigr),
\label{eq:kappa}
\end{eqnarray}
where $\kappa_0\equiv\frac{2\pi^2\tau T^3}{15c}$ is the normal phonon thermal conductivity. $\Delta_j$ is the splitting of Pr doublet at site $j$ due to the ``local effective field", i.e. the combined effects of external magnetic field and the exchange interaction with surrounding doublets. $\delta$ is the variance of local effective field, which is essential to the resonant spin-phonon scattering, and is usually attributed to the randomness in the system.

In the present analysis, we adopt the nearest-neighbor spin ice model to describe the thermal fluctuation of Pr doublets, 
\begin{eqnarray}
\mathcal{H} = J\sum_{\langle j,j'\rangle}\sigma_j\sigma_{j'} - {\mathbf h}\cdot\sum_j\sigma_j{\mathbf d}_j.
\end{eqnarray}
Here the first term is the nearest-neighbor interaction between Pr doublets, $\sigma_j=\pm1$. The second term describes the site-dependent Zeeman interaction with external magnetic field, ${\mathbf h}$.
${\mathbf d}_j$ stands for the easy axis of the Pr doublet at site $j$. We assume the case of $[001]$ field direction, and conducted the Monte Carlo simulation for $N=16\times16\times16\times4=16384$ doublets and, made 10000 samplings for the effective field, $\Delta_j\equiv(\frac{2}{\sqrt{3}}h - 2J\sum_{j'}\sigma_{j'})\sigma_j$.
From the thermal average, we obtain the thermal conductivity through Eq.~(\ref{eq:kappa}).

\clearpage
\begin{center}
\includegraphics[width=1.0\textwidth]{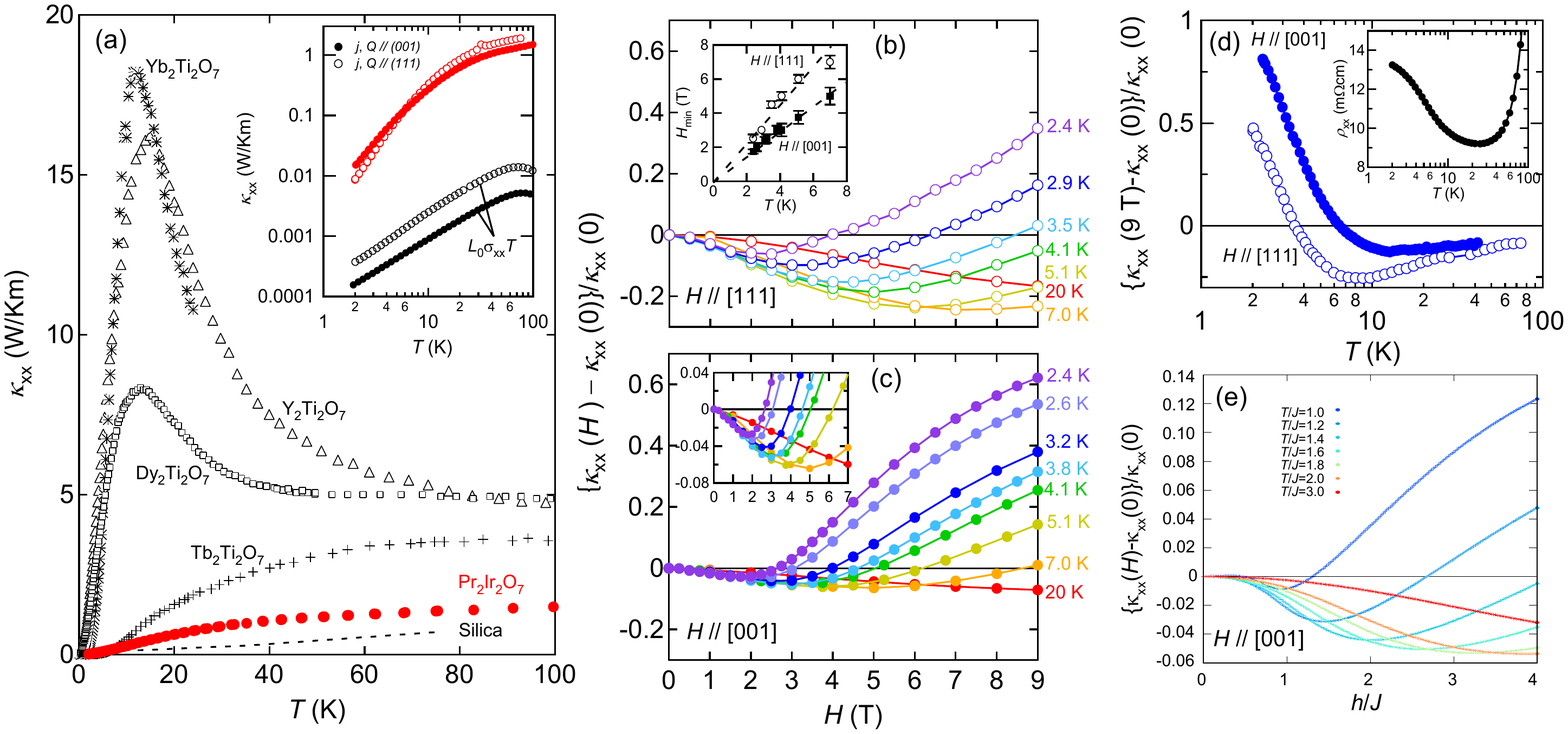}
\end{center}
\vspace{-1.0cm}
FIG.1. (a) Temperature dependence of zero-field longitudinal thermal conductivity $\kappa_{xx}$ for the heat current $Q$ parallel to the (001) plane together with those of the pyrochlore compounds~\cite{tokiwa,kolland,li}. Inset shows a $\kappa_{xx}$ vs $T$ plot in a logarithmic scale for $Q\parallel$ (001) and $Q\parallel$ (111). The electronic contribution $L_0\sigma_{xx}T$ in $\kappa_{xx}$ estimated by using the Wiedemann-Franz law is also shown for the electrical current $j$ parallel to the (001) and (111) planes. Magnetic field dependence of longitudinal thermal conductivity normalized by the zero-field value $\{\kappa_{xx}(H)-\kappa_{xx}(0)\}/\kappa_{xx}(0)$ at different temperatures under the magnetic fields parallel to the [111] (b) and [001] (c) directions. Inset of panel (b) depicts a $H_{\rm min}$ vs $T$ plot for $H\parallel[111]$ and $H\parallel[001]$. A zoom of the low field region for the $H\parallel[001]$ data is shown in the inset of panel (c). (d) Temperature dependence of $\{\kappa_{xx}(H)-\kappa_{xx}(0)\}/\kappa_{xx}(0)$ at $H$ = 9 T for $H\parallel[111]$ and $H\parallel[001]$. Temperature dependence of the longitudinal electrical resistivity $\rho_{xx}$ at zero field is shown in the inset. (e) The calculated $\{\kappa_{xx}(H)-\kappa_{xx}(0)\}/\kappa_{xx}(0)$ as a function of $h/J$ with various values of $T/J$ for $H\parallel[001]$, where $h$ is external magnetic field and $J$ is nearest-neighbor interaction between Pr doublets.

\clearpage
\begin{center}
\includegraphics[width=0.8\textwidth]{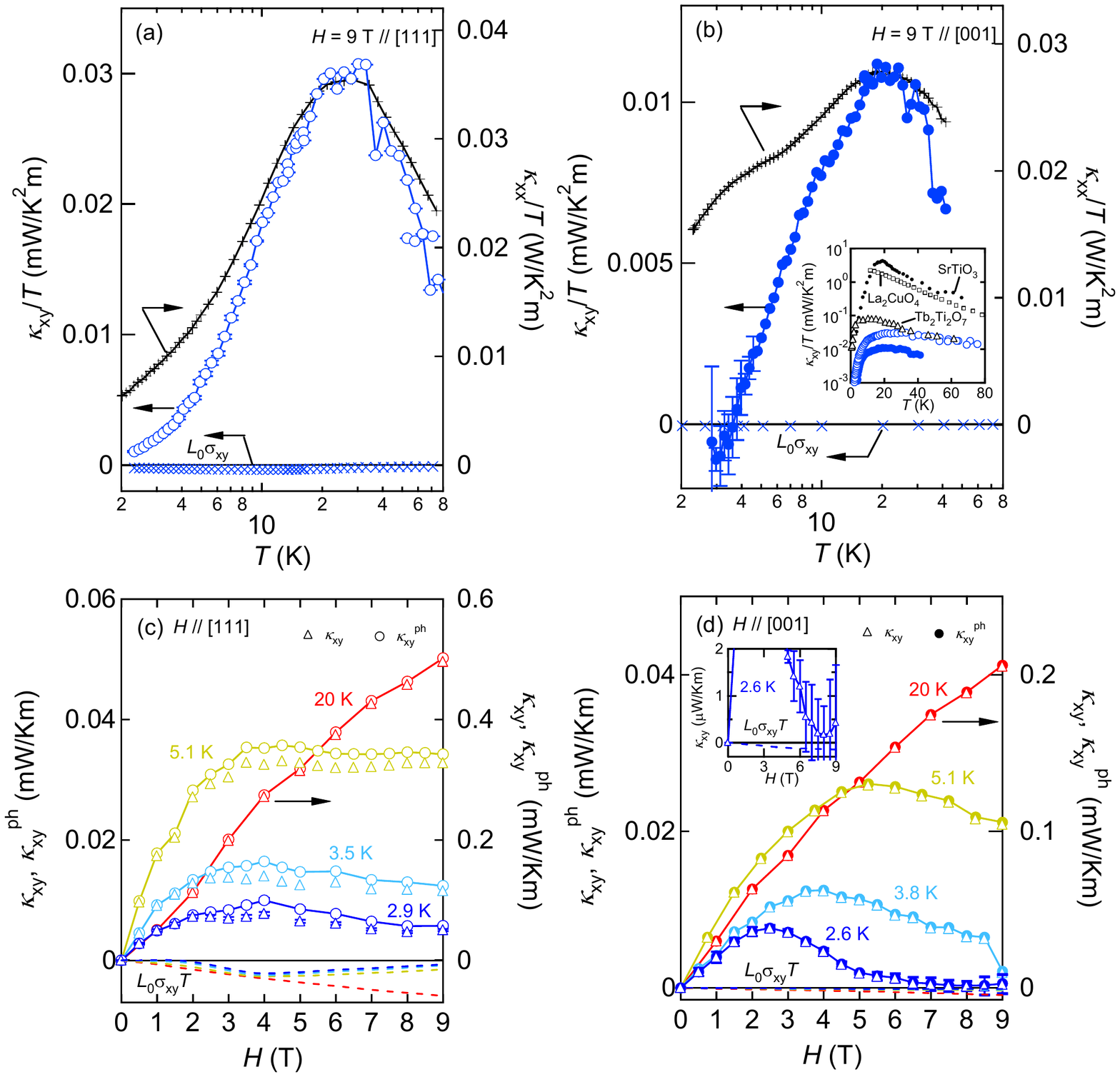}
\end{center}
\vspace{-5cm}
FIG.2. Temperature dependence of thermal Hall conductivity divided by temperature $\kappa_{xy}/T$ and $L_0\sigma_{xy}$ (left axis) together with longitudinal thermal conductivity divided by temperature $\kappa_{xx}/T$ (right axis) under the magnetic field of 9 T applied parallel to the [111] (a) and [001] (b) directions. In the inset of panel (b), our data is compared with those of Tb$_2$Ti$_2$O$_7$~\cite{hirschberger}, SrTiO$_3$~\cite{STO}, and cuprate Mott insulator~\cite{grissonnanche1}. Magnetic field dependence of $\kappa_{xy}$ (triangles) and $L_0\sigma_{xy}T$ (dotted lines) at different temperatures for $H\parallel[111]$ (c) and $H\parallel[001]$ (d). Phonon contribution estimated by $\kappa_{xy}^{ph}=\kappa_{xy}-L_0\sigma_{xy}T$ is also shown by circles. $\kappa_{xy}$ for $H\parallel[001]$ seems to approach the $L_0\sigma_{xy}T$ value at high fields as displayed in the inset of panel (d).

\clearpage
\begin{center}
\includegraphics[width=1\textwidth]{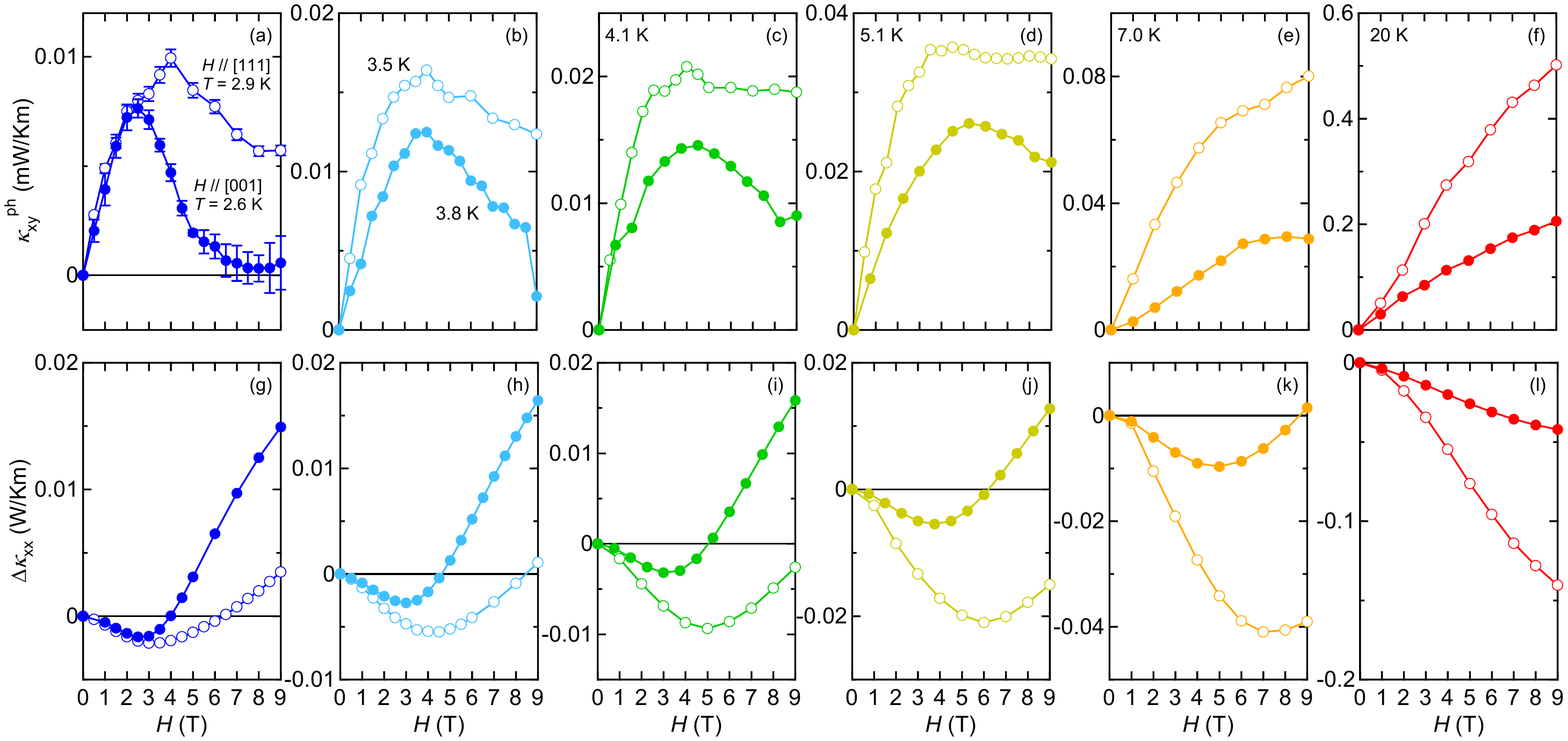}
\end{center}
\vspace{-2cm}
FIG.3. Magnetic field dependence of thermal Hall conductivity of phonons $\kappa_{xy}^{ph}$ (upper panels) and magnetic field-induced change in the longitudinal thermal conductivity $\Delta\kappa_{xx}=\kappa_{xx}(H)-\kappa_{xx}(0)$ (lower panels) for $H\parallel[111]$ (open circles) and $H\parallel[001]$ (closed circles). Measurements are performed at (a, g) $T=$ 2.9 K and 2.6 K for $H\parallel[111]$ and $H\parallel[001]$, respectively, (b, h) $T=$ 3.5 K and 3.8 K for $H\parallel[111]$ and $H\parallel[001]$, respectively, (c, i) 4.1 K, (d, j) 5.1 K, (e, k) 7.0 K, and (f, l) 20 K.

\clearpage

\begin{center}
\includegraphics[width=0.8\textwidth]{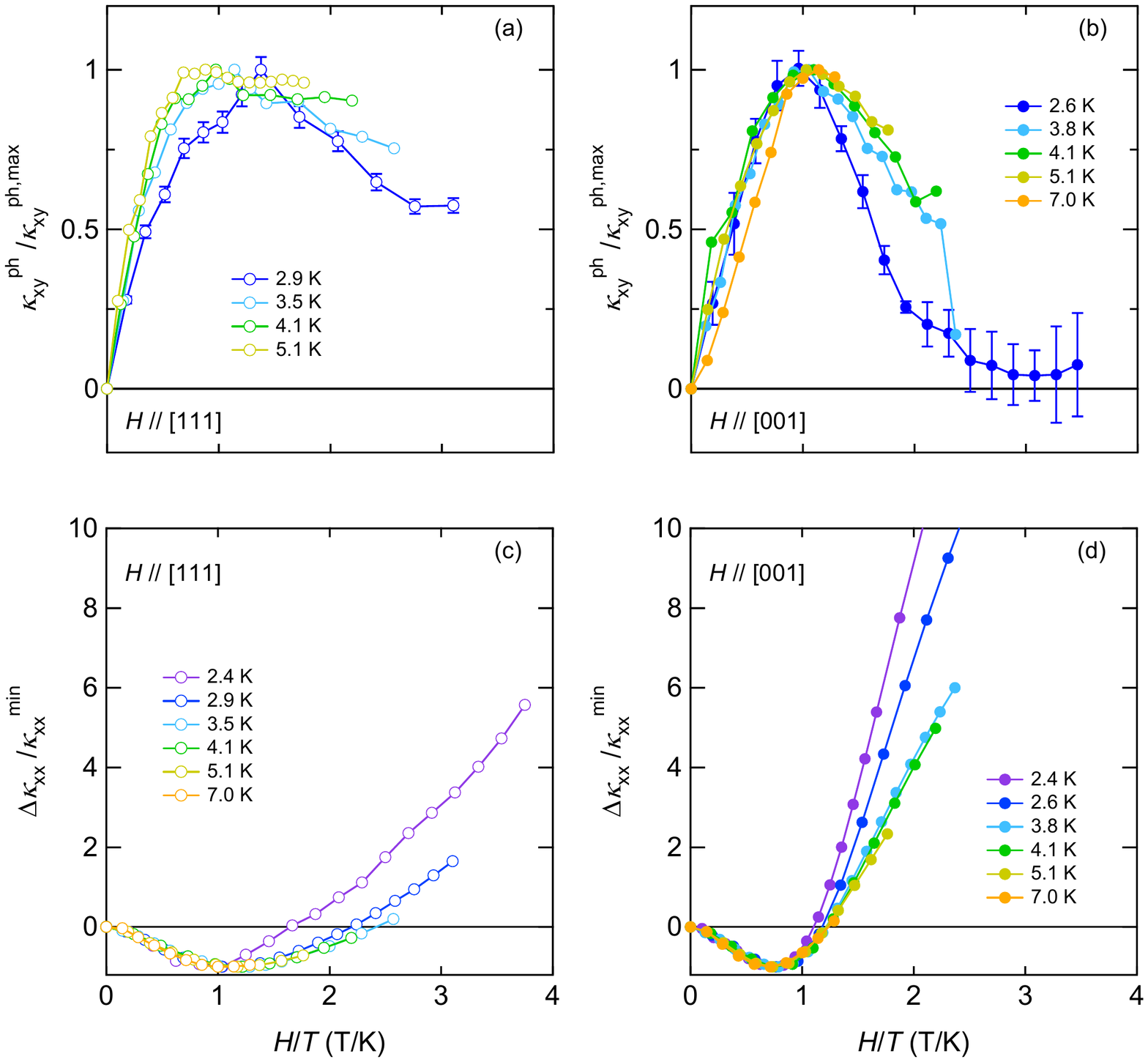}
\end{center}
\vspace{-6cm}
FIG.4. Phonon thermal Hall conductivity normalized by the maximum value $\kappa_{xy}^{ph}/\kappa_{xy}^{ph,\rm max}$ as a function of $H/T$ for $H\parallel[111]$ (a) and $H\parallel[001]$ (b). Magnetic field-induced change in the longitudinal thermal conductivity normalized by the minimum value $\Delta\kappa_{xx}/\kappa_{xx}^{\rm min}$ as a function of $H/T$ for $H\parallel[111]$ (c) and $H\parallel[001]$ (d).

\end{document}